\documentclass[runningheads]{llncs}

\newif\ifappendix
\appendixfalse

\usepackage[numbers]{natbib}

\usepackage[T1]{fontenc}
\usepackage{enumitem}
\usepackage{booktabs}
\usepackage{hyperref}
\usepackage{tikz}
\usetikzlibrary{myautomata}
\usetikzlibrary{decorations.pathreplacing,calc}
\usepackage{caption}
\usepackage{subcaption}
\usepackage{multicol}
\usepackage{accsupp}
\usepackage{amssymb} 
\usepackage{mathtools} 
\usepackage[vlined]{algorithm2e}
\providecommand*{\napprox}{%
  \BeginAccSupp{method=hex,unicode,ActualText=2249}%
  \not\approx
  \EndAccSupp{}%
}
\newcommand{\varof}{\mathbin{\approx}}
\newcommand{\nvarof}{\mathbin{\napprox}}
\newcommand{\specof}{\mathbin{\sqsubseteq}}
\newcommand{\tikzmark}[1]{\tikz[overlay,remember picture] \node (#1) {};}

\definecolor{lime}{HTML}{A6CE39}
\DeclareRobustCommand{\orcidicon}{
	\hspace{-2mm}
	\begin{tikzpicture}
	\draw[lime, fill=lime] (0,0)
	circle [radius=0.16]
	node[white] (ID) {{\fontfamily{qag}\selectfont \tiny ID}};
	\draw[white, fill=white] (-0.0625,0.095)
	circle [radius=0.007];
	\end{tikzpicture}
	\hspace{-2.5mm}
      }
\def\orcidID#1{\href{https://orcid.org/#1}{\smash{\orcidicon}}}

\title{Effective Reductions of Mealy Machines}

\author{Florian Renkin\orcidID{0000-0002-5066-1726} \and Philipp Schlehuber-Caissier\orcidID{0000-0002-6611-9659} \and Alexandre Duret-Lutz\orcidID{0000-0002-6623-2512} \and Adrien Pommellet\orcidID{0000-0001-5530-152X}}

\institute{
  LRDE, EPITA, Kremlin-Bicêtre, France  \email{\{frenkin,philipp,adl,adrien\}@lrde.epita.fr}\\
}

\authorrunning{F. Renkin \and P. Schlehuber-Caissier \and A. Duret-Lutz \and A. Pommellet}

\usetikzlibrary{automata}
\usetikzlibrary{arrows.meta}
\usetikzlibrary{bending}
\usetikzlibrary{quotes}
\usetikzlibrary{positioning}
\usetikzlibrary{calc}

\tikzset{
  automaton/.style={
    semithick,shorten >=1pt,>={Stealth[round,bend]},
    node distance=1cm,
    initial text=,
    every initial by arrow/.style={every node/.style={inner sep=0pt}},
    every state/.style={minimum size=7.5mm,fill=white}
  },
  smallautomaton/.style={
    automaton,
    node distance=5mm,
    every state/.style={minimum size=4mm,fill=white,inner sep=1pt}
  },
  mediumautomaton/.style={
    automaton,
    node distance=1.5cm,
    every state/.style={minimum size=6mm,fill=white,inner sep=1pt}
  },
  initial overlay/.style={every initial by arrow/.style={overlay}},
  accset/.style={
    fill=blue!50!black,draw=white,text=white,thin,
    circle,inner sep=1pt,anchor=center,font=\bfseries\sffamily\tiny
  },
  color acc0/.style={fill=magenta},
  color acc1/.style={fill=cyan},
  color acc2/.style={fill=orange},
  color acc3/.style={fill=green!70!black},
  color acc4/.style={fill=blue!50!black},
}

\makeatletter
\tikzoption{initial angle}{\tikzaddafternodepathoption{\def\tikz@initial@angle{#1}}}
\makeatother

\tikzstyle{initial overlay}=[every initial by arrow/.style={overlay}]

\tikzstyle{state-labels}=[state/.style=state with output,inner sep=2pt]

\tikzstyle{statename}=[
below,label distance=2pt,
fill=yellow!30!white,
rounded corners=1mm,inner sep=2pt
]

\tikzstyle{accset}=[
fill=blue!50!black,draw=white,text=white,thin,
circle,inner sep=.9pt,anchor=center,font=\bfseries\sffamily\tiny
]

\tikzset{
  ks/.style={},
  collacc0/.style={fill=blue!50!cyan},
  collacc1/.style={fill=magenta},
  collacc2/.style={fill=orange!90!black},
  collacc3/.style={fill=green!70!black},
  collacc4/.style={fill=blue!50!black},
  fs/.style={font=\bfseries\sffamily\small},
  acc/.pic={\node[text width={},text height={},minimum size={0pt},accset,collacc#1,ks]{#1};},
  accs/.pic={\node[text width={},text height={},minimum size={0pt},accset,collacc#1,fs,ks]{#1};},
  starnew/.pic={\node[text width={},text height={},minimum size={0pt},text=magenta]{$\filledlargestar$};},
  starimpr/.pic={\node[text width={},text height={},minimum size={0pt},text=blue!50!cyan]{$\filledlargestar$};},
  balldigit/.style={text=white,circle,minimum size={12pt},shade,ball color=structure.fg,inner sep=0pt,font={\footnotesize\bf},align=center}
}

\makeatletter
\def\tikzopacityregister{1}
\tikzset{
  opacity/.append code={
    \pgfmathsetmacro\tikzopacityregister{#1*\tikzopacityregister}
  },
  opacity aux/.code={ 
    \tikz@addoption{\pgfsetstrokeopacity{#1}\pgfsetfillopacity{#1}}
  },
  every shadow/.style={opacity aux=\tikzopacityregister},
  covered/.style={opacity=0},
  uncover on/.style={alt={#1{}{covered}}},
  alt/.code args={<#1>#2#3}{%
    \alt<#1>{\pgfkeysalso{#2}}{\pgfkeysalso{#3}} 
  },
  explains/.style={rectangle callout,callout absolute pointer={#1},fill=structure.fg!10,drop shadow={fill=black!70!structure.fg!30},align=center}
}
\makeatother

\newcommand{\B}{\ensuremath{\mathbb{B}}}
\newcommand{\K}{\ensuremath{\mathbb{K}}}
\newcommand{\Succ}{\ensuremath{\mathrm{Succ}}}
\newcommand{\Out}{\ensuremath{\mathrm{Out}}}

\newcommand{\tup}[1]{{\ensuremath{\left(#1\right)}}}
\newcommand{\set}[1]{{\ensuremath{\left\lbrace#1\right\rbrace}}}

\tikzset{
  automaton/.style={
    semithick,shorten >=1pt,
    node distance=1.5cm,
    initial text=,
    every initial by arrow/.style={every node/.style={inner sep=0pt}},
    every state/.style={
      align=center,
      fill=white,
      minimum size=7.5mm,
      inner sep=0pt,
      execute at begin node=\strut,
    }},
  smallautomaton/.style={automaton,
                         node distance=7mm,
                         every state/.style={minimum size=4mm,
                           fill=white,
                           inner sep=1.5pt}},
  >={Stealth[round,bend]},
}

\begin{document}
\maketitle

\begin{abstract}
  We revisit the problem of reducing incompletely specified Mealy
  machines with reactive synthesis in mind.  We propose two
  techniques: the former is inspired by the tool {\sc
    MeMin}~\citet{abel.15.iccad} and solves the minimization problem,
  the latter is a novel approach derived from simulation-based
  reductions but may not guarantee a minimized machine.  However, we
  argue that it offers a good enough compromise between the size of
  the resulting Mealy machine and performance.  The proposed methods
  are benchmarked against \textsc{MeMin} on a large collection of test
  cases made of well-known instances as well as new ones.\\
\end{abstract}

\section{Introduction}

\begin{figure}[b]
\begin{subfigure}[t]{0.28\textwidth}
{\centering
  \begin{tikzpicture}[mediumautomaton,node distance=1.1cm and 2.1cm]
    \node[draw,minimum width=1.4cm,minimum height=1.4cm] (C) {};
    \draw[->] (C.160) +(-5mm,0) node[left]{$a$} -- (C.160);
    \draw[->] (C.-160) +(-5mm,0) node[left]{$b$} -- (C.-160);
    \draw[->] (C.20) -- ++(5mm,0) node[right]{$x$};
    \draw[->] (C.-20) -- ++(5mm,0) node[right]{$y$};
  \end{tikzpicture}
  \caption{A reactive controller}
  \label{controler}}
\end{subfigure}
\hfill
\begin{subfigure}[t]{0.39\textwidth}
{\centering
  \begin{tikzpicture}[mediumautomaton,node distance=1cm and 1.33cm]
    \node[initial,initial angle=90, lstate] (v0) {$0$};
    \node[lstate,right=of v0] (v1) {$1$};
    \node[lstate,left=of v0] (v2) {$2$};
    \path[->]
    (v0) edge[bend left=14] node[above,align=center] {$ab/\{x\bar{y},\mathrlap{xy\}}$\\$a\bar{b}/\set{\bar{x}y}$} (v1)
    (v0) edge node[above] {$\bar{a}\bar{b}/\set{\bar{x}\bar{y}}$} (v2)
    (v1) edge[bend left=14] node[below,align=center] {$\bar{a}\bar{b}/\set{x\bar{y},\bar{x}\bar{y}}$\\$ab/\set{x\bar{y}}$} (v0)
        (v2) edge[loop below,looseness=10] node[right=2pt] {$\bar{a}\bar{b}/\set{\bar{x}\bar{y}}$} (v2)
    ;
  \end{tikzpicture}
  \caption{Original machine}
  \label{autEx1}}
\end{subfigure}
\hfill
\begin{subfigure}[t]{0.24\textwidth}
 {\centering
  \begin{tikzpicture}[mediumautomaton,node distance=1.7cm and 2.1cm]
    \begin{scope}[local bounding box=aut]
      \node[initial,lstate] (v0) {$0$};
    \end{scope}
    \path[->]
      (v0) edge[loop above,looseness=10] node[right=2pt] {$ab/\set{x\bar{y}}$} (v0)
    (v0) edge[loop right,looseness=10] node {$a\bar{b}/\set{\bar{x}y}$} (v0)
    (v0) edge[loop below,looseness=10] node[right=2pt] {$\bar{a}\bar{b}/\set{\bar{x}\bar{y}}$} (v0)
    ;
  \end{tikzpicture}
  \caption{Minimal machine}
  \label{autEx1_Min}}
\end{subfigure}
  \caption{Minimizing a Mealy machine that models a reactive controller}
  \label{autEx1_all}
\end{figure}
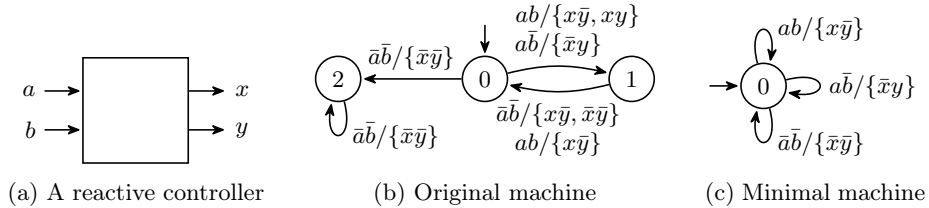

Program synthesis is a well-established formal method: given a logical
specification of a system, it allows one to automatically generate a
provably correct implementation. It can be applied to reactive
controllers (Fig.~\ref{controler}): circuits that produce for an input
stream of Boolean valuations (here, over Boolean variables $a$ and
$b$) a matching output stream (here, over $x$ and $y$).

The techniques used to translate a specification (say, a Linear Time
Logic formula that relates input and output Boolean variables) into a
circuit often rely on automata-theoretic intermediate models such as
Mealy machines.  These transducers are labeled graphs whose edges
associate input valuations to a choice of one or more output
valuations, as shown in Fig.~\ref{autEx1}.

Since Mealy machines with fewer states result in smaller circuits,
reducing and minimizing the size of Mealy machines are
well-studied problems~\citet{alberto.09.latw, paull.59.tec}.

However, vague specifications may cause incompletely specified
machines: for some states (i.e., nodes of the graph) and inputs,
there may not exist a unique, explicitly defined output, but
a set of valid outputs.  Resolving those choices to a single output
(among those allowed) will produce a fully specified machine that
satisfies the initial specification, however those different choices
may have an impact on the minimization of the machine.  While
minimizing fully specified machines is efficiently
solvable~\citet{hopcroft.71.tmc}, the problem is NP-complete for
incompletely specified machines~\citet{pfleeger.73.tc}. Hence, it may
also be worth exploring faster algorithms that seek to reduce the
number of states without achieving the optimal result.

Consider Fig. \ref{autEx1}: this machine is incompletely specified, as
for instance state $0$ allows multiple outputs for input $ab$
(i.e., when both input variables $a$ and $b$ are true) and implicitly
allows any output for input $\bar a b$ (i.e., only $b$ is true) as it isn't
constrained in any way by the specification.  We
can benefit from this flexibility in unspecified outputs to help
reduce the automaton.  For instance if we constrain state 2 to behave
exactly as state 0 for inputs $ab$ and $a \bar b$, then these two
states can be merged.  Adding further constraints can lead to the
single-state machine shown in Fig. \ref{autEx1_Min}.  These smaller
machines are not \emph{equivalent}, but they are \emph{compatible}: for
any input stream, they can only produce output streams that could also
have been produced by the original machine. 

We properly define \emph{Incompletely specified Generalized Mealy
  Machines} in Section \ref{secDef} and provide a SAT-based
minimization algorithm in Section \ref{secMin}.
Since the minimization of incompletely specified Mealy machines is desirable but
not crucial for reactive synthesis, we propose a faster reduction
technique yielding ``small enough'' machines in Section
\ref{secBisim}.  Finally, in Section \ref{secBench} we benchmark these
techniques against the state-of-the-art tool {\sc
  MeMin}~\citet{abel.15.iccad}.




\section{Definitions}\label{secDef}

Given a set of propositions (i.e., Boolean variables) $X$,
let $\B^X$ be the set of all possible valuations on $X$, and
let $2^{\B^X}$ be its set of subsets.
Any element of $2^{\B^X}$ can be expressed as a Boolean formula over $X$.
The negation of proposition $p$ is denoted $\bar{p}$.
We use $\top$ to denote the Boolean formula that is always true, or
equivalently the set $\B^X$, and assume that $X$ is clear from the context.
A \emph{cube} is a conjunction of propositions or their negations (i.e., literals).
As an example, given three propositions $a$, $b$ and $c$,
the cube $a \land \bar{b}$, written $a\bar{b}$,
stands for the set of all valuations such that $a$ is true and $b$ is false,
i.e. $\{a\bar{b}c, a\bar{b}\bar{c}\}$.
Let $\K^X$ stand for the set of all cubes over $X$.
$\K^X$ contains the cube $\top$, that stands
for the set of all possible valuations over $X$.
Note that any set of valuations can be represented
as a disjunction of disjoint cubes (i.e., not sharing a common valuation).


\begin{definition}
  An \emph{Incompletely specified Generalized Mealy Machine} (IGMM) is
  a tuple $M=\tup{I, O, Q, q_{\mathit{init}}, \delta, \lambda}$, where
  $I$ is a set of \emph{input propositions},
  $O$ a set of \emph{output propositions},
  $Q$ a finite set of \emph{states}, $q_{\mathit{init}}$ an \emph{initial state},
  $\delta \colon \left(Q, \B^{I}\right) \rightarrow Q$
  a partial \emph{transition function}, and
  $\lambda \colon \left(Q, \B^{I}\right) \rightarrow 2^{\B^{O}}\setminus \{\emptyset\}$
  an \emph{output function} such that $\lambda(q,i)=\top$ when $\delta(q,i)$ is undefined.
  If $\delta$ is a total function, we then say that $M$ is \emph{input-complete}.
\end{definition}

It is worth noting that the transition function is input-deterministic
but not complete with regards to $Q$ as $\delta(q,i)$ could be
undefined.  Furthermore, the output function may return many valuations
for a given input valuation and state.  This is not an unexpected
definition from a reactive synthesis point of view, as a given
specification may yield multiple compatible output valuations for a
given input.



\begin{definition}[Semantics of IGMMs]
	Let $M=\tup{I, O, Q, q_{\mathit{init}}, \delta, \lambda}$ be an
	IGMM. For all $u \in \B^{I}$ and $q \in Q$, if $\delta(q, u)$ is defined,
	we write that $q \xrightarrow{u / v} \delta(q, u)$ for all
	$v \in \lambda(q, u)$. Given two infinite sequences of valuations
	$\iota=i_0\cdot i_1\cdot i_2\cdots\in (\B^{I})^\omega$ and
	$o=o_0\cdot o_1\cdot o_2\cdots\in (\B^{O})^\omega$,
	$(\iota,o)\models M_q$ if and only if:
	\begin{itemize}
		\item either there is an infinite sequence of states
		$(q_j)_{j \ge 0} \in Q^\omega$ such that $q = q_0$ and
		$q_0 \xrightarrow{i_0 / o_0} q_1 \xrightarrow{i_1 / o_1} q_2
		\xrightarrow{i_2 / o_2} \cdots$;
		\item or there is a finite sequence of states
		$(q_j)_{0 \le j \le k} \in Q^{k+1}$ such that $q = q_0$, $\delta(q_k, i_k)$ is
		undefined, and
		$q_0 \xrightarrow{i_0 / o_0} q_1 \xrightarrow{i_1 / o_1} \cdots q_k$.
	\end{itemize}
	We then say that starting from state $q$, $M$ produces output $o$
	given the input $\iota$.
\end{definition}

Note that if $\delta(q_k,i_k)$ is undefined, the machine is allowed to
produce an arbitrary output from then on.  Furthermore, given an input
word $\iota$, there may be several output words $o$ such that
$(\iota,o) \models M_q$ (in accordance with a lax specification).

As an example, consider the input sequence
$\iota = ab\cdot \bar a\bar b\cdot  ab\cdot \bar a\bar b\cdots$
applied to the initial state $0$ of the machine shown in Figure~\ref{autEx1}.
We have $(\iota,o)\models M_0$ if and only if for all $j \in \mathbb{N}$,
$o_{2j}\in x$ and $o_{2j+1}\in \bar y$, where $x$ and $\bar y$ are
cubes that respectively represent $\{xy,x\bar y\}$ and $\{x\bar y,\bar x\bar y\}$.

\begin{definition}[Variation and specialization]
  Let $M=\tup{I, O, Q, q_{\mathit{init}}, \delta, \lambda}$ and
  $M'=\tup{I, O, Q', q'_{\mathit{init}}, \delta', \lambda'}$ be two IGMMs.
  Given two states $q \in Q$, $q' \in Q'$, we say that $q'$ is a:
  \begin{itemize}[noitemsep,topsep=2pt]
    \item \emph{variation} of $q$ if $\forall \iota \in
    ({\B^I})^\omega ,
    \set{o \mid (\iota,o) \models M'_{q'}} \cap \set{o \mid (\iota,o)\models M_{q}}
    \neq \emptyset$;
    \item \emph{specialization} of $q$ if $\forall \iota \in
    ({\B^I})^\omega ,
    \set{o \mid (\iota,o) \models M'_{q'}} \subseteq \set{o \mid (\iota,o)\models M_{q}}$.
  \end{itemize}
  We say that $M'$ is a variation (resp.\ specialization) of $M$
  if $q_{\mathit{init}}'$ is a variation (resp.\ specialization)
  of $q_{\mathit{init}}$.
\end{definition}

Intuitively, all the input-output pairs accepted by
a specialization $q'$ in $M'$ are also accepted by $q$ in $M$.
Therefore, if all the outputs produced by state $q$ in $M$
comply with the original specification, then so do the outputs produced
by state $q'$ in $M'$.
In order for two states to be a variation of one another,
for all possible inputs they must be able to agree on a common output behaviour.

We write $q'\varof{} q$ (resp. $q' \specof{} q$) if $q'$ is a
variation (resp. specialization) of $q$.  Note that $\varof{}$ is a
symmetric but non-transitive relation, while $\specof$ is transitive
($\specof$ is a preorder).

\medskip
Our goal in this article is to solve the following problems:
\begin{description}[noitemsep,topsep=2pt]
  \item[Reducing an IGMM $M$:] finding a specialization of $M$ having at most
  the same number of states, preferably fewer.
  \item[Minimizing an IGMM $M$:] finding a specialization of $M$ having the
  least number of states.
\end{description}

Consider again the IGMM shown in Figure~\ref{autEx1}.
The IGMM shown in Figure~\ref{autEx1_Min} is a specialization of this machine
and has a minimal number of states.

\subsubsection*{Generalizing inputs and outputs.}
\label{secCompToReg}

Note that the output function of an IGMM returns a set of valuations,
but it can be rewritten equivalently to output a set of cubes
as $\lambda \colon \left(Q, \B^I\right) \rightarrow 2^{\K^{O}}$.
As an example, consider $I = \{a\}$ and $O = \{x, y, z\}$; the set of valuations
$v = \{\bar{x}yz, \bar{x}y\bar{z}, x\bar{y}z, x\bar{y}\bar{z}\}\in 2^{\B^O}$ is equivalent to the
set of cubes $v_c = \{\bar{x}y, x\bar{y}\}\in 2^{\K^O}$. 

In the literature, a Mealy machine commonly maps a single input
valuation to a single output valuation: its output function is therefore of the
form $\lambda \colon \left(Q, \B^I\right) \rightarrow \B^{O}$.  The
tool \textsc{MeMin}~\citet{abel.15.iccad} uses a slight generalization by allowing a
single output cube, hence $\lambda \colon \left(Q, \B^{I}\right) \rightarrow \K^{O}$. Thus,
unlike our model, neither the common definition nor the tool \textsc{MeMin} can
feature an edge outputting the aforementioned set $v$ (or equivalently $v_c$),
as it cannot be represented by a single cube or valuation.
Our model is therefore \emph{strictly more expressive}, although
it comes at a price for minimization.

Note that, in practice, edges with identical source state,
output valuations, and destination state can be merged into a single transition
labeled by the set of allowed inputs. Both our tool and \textsc{MeMin} feature
this optimization. While it does not change the
expressiveness of the underlying model, this more succinct representation
of the machines does improve the efficiency of the algorithms
detailed in the next section, as they depend on the total number of transitions.

\section{SAT-Based Minimization of IGMM}
\label{secMin}

This section builds upon the approach presented
by~\citet{abel.15.iccad} for machines with outputs constrained to
cubes, and generalizes it to the IGMM model (with more
expressive outputs).

\subsection{General approach}

\begin{definition}
	Given an IGMM $M=\tup{I, O, Q, q_{\mathit{init}}, \delta, \lambda}$,
	\emph{a variation class} $C \subseteq Q$ is a set of states such that all
	elements are pairwise variations, i.e. $\forall q,q' \in C$, $q'\varof{} q$.
	For any input $i \in \B^I$, we define:
	\begin{itemize}[noitemsep,topsep=2pt]
		\item the \emph{successor function}
		$\Succ(C, i) =
      \bigcup_{q\in C} \set{\delta(q,i) \mid
			\delta(q,i) \text{~is defined}} $;
		\item the \emph{output function}
		$\Out(C, i) = \bigcap_{q\in C} \lambda(q,i)$.
	\end{itemize}
\end{definition}

Intuitively, the successor function returns the set of all states
reachable from a given class under a given input symbol.
The output function returns the set of all shared output valuations
between the various states in the class.

In the remainder of this section we will call a variation class simply a class,
as there is no ambiguity. We consider three important notions concerning
classes, or rather sets thereof, of the form $S = \set{C_0,\ldots,C_{n-1}}$.

\begin{definition}[Cover condition]\label{cond_cover}
	We say that a set of classes $S$ \emph{covers} the machine $M$ if every
	state of $M$ appears in at least one of the classes.
\end{definition}

\begin{definition}[Closure condition]\label{cond_closure}
  We say that a set of classes $S$ is \emph{closed} if for all
  $C_j\in S$ and for all inputs $i \in \B^I$ there exists a $C_k\in S$
  such that $\Succ(C_j, i) \subseteq C_k$.
\end{definition}

\begin{definition}[Nonemptiness condition]\label{cond_nonempt}
	We say that a class $C$ has a \break\emph{nonempty output} if $\Out(C,i)\ne\emptyset$ for
	all inputs $i\in \B^I$.
\end{definition}

The astute reader might have observed that the nonempty output condition
is strictly stronger than the condition that all elements in a class
have to be pairwise variations of one another.
We will see that this distinction is however important, as it gives rise to a
different set of clauses in the SAT problem, reducing the total runtime.

Combining these conditions yields the main theorem for this approach.
This extends a similar theorem by~\citet[][Thm~1]{abel.15.iccad} by
adding the nonemptiness condition to support the more expressive IGMM
model.

\begin{theorem}
  Let $M=\tup{I, O, Q, q_{\mathit{init}}, \delta, \lambda}$ be an IGMM and
  $S = \set{C_0,\ldots,C_{n-1}}$ be a \emph{minimal} (in terms of size)
  set of classes such that \emph{(1)} $S$ is \emph{closed},
  \emph{(2)} $S$ \emph{covers} every state of the machine $M$ and
  \emph{(3)} each of the classes $C_j$ has a \emph{nonempty output}.
  Then the IGMM $M'=\tup{I,  O, S, q_{\mathit{init}}', \delta', \lambda'}$ where:
  \begin{itemize}[noitemsep,topsep=2pt]
	 \item $q_{\mathit{init}}' = C$ for some $C \in S$ such that
	 $q_{\mathit{init}} \in C$;
	 \item $\delta'(C_j, i) = \begin{cases}C_k \text{ for some k s.t. } \Succ(C_j, i) \subseteq C_k &\text{if~} \Succ(C_j, i) \neq \emptyset \\
                                         \text{undefined} &\text{else;}
                            \end{cases}$
   \item $\lambda'(C_j, i) = \begin{cases} \Out(C_j, i) &\text{if~} \Succ(C_j, i) \neq \emptyset\\
                                           \top &\text{else;}
                             \end{cases}$
  \end{itemize}
  is a \emph{specialization} of minimal size (in terms of states) of $M$.
  \label{theoremSATMIN}
\end{theorem}

\begin{figure}[t]
	\begin{subfigure}[t]{0.59\textwidth}
		\centering
		\begin{tikzpicture}[mediumautomaton,node distance=1cm and 1.186cm]
      \begin{scope}[local bounding box=aut]
        \node[initial,lstate, fill = violet!50] (v0) {$0$};
        \node[lstate,above=of v0, fill = orange!50] (v1) {$1$};
        \node[lstate,right=of v0, fill = green!50] (v2) {$2$};
        \node[lstate,right=of v1, fill = orange!50] (v3) {$3$};
        \node[lstate,right=of v2, fill = green!50] (v5) {$5$};
        \node[lstate,right=of v5, fill = orange!50] (v4) {$4$};
                          \fill[fill=green!50] (v4.center) -- (v4.east) arc (0:120:2.99mm) -- cycle;
                          \fill[fill=violet!50] (v4.center) -- (v4.east) arc (0:-120:2.99mm) -- cycle;
        \node[lstate,fill=none] at (v4) {$4$};
        \node[lstate,right=of v3, fill = orange!50] (v6) {};
                          \fill[fill=green!50] (v6.center) -- (v6.east) arc (0:120:2.99mm) -- cycle;
                          \fill[fill=violet!50] (v6.center) -- (v6.east) arc (0:-120:2.99mm) -- cycle;
        \node[lstate,fill=none] at (v6) {$6$};
      \end{scope}
      \path[->]
      (v0) edge node[left] {$a/{\bar{z}}$} (v1)
      (v0) edge node[above] {$\bar{a}/\bar{x}\bar{y}\bar{z}$} (v2)
      (v1) edge[loop left] node {$a/{\bar{z}}$} (v1)
      (v1) edge[bend left=10, above] node {$\bar{a}/{z}$} (v3)
      (v2) edge[bend right=20] node[below] {$a/\top$} (v4)
      (v2) edge[above] node {$\bar{a}/{z}$} (v5)
      (v3) edge[bend left=10, below] node {$a/{\bar{z}}$} (v1)
      (v3) edge[above] node {$\bar{a}/\top$} (v6)
      (v5) edge[above] node {$\bar{a}/\top$} (v4)
      (v4) edge[loop above] node[align=center] {$a/\top$\\$\bar{a}/\top$} (v4)
      (v5) edge[loop above] node {$a/{z}$} (v5)
      (v6) edge[loop right] node[align=center] {$a/\top$\\$\bar{a}/\top$} (v6)
      ;
		\end{tikzpicture}
		\subcaption{Original IGMM $M$}
		\label{autExSatBase}
	\end{subfigure}
	\begin{subfigure}[t]{0.4\textwidth}
		\centering
		\begin{tikzpicture}[mediumautomaton,node distance=1.cm and 1.4cm]
			\begin{scope}[local bounding box=aut]
				\node[initial,lstate, fill = violet!50] (v0) {$0$};
				\node[lstate,above= of v0, fill = orange!50] (v1) {$1$};
				\node[lstate,right= of v0, fill = green!50] (v2) {$2$};
			\end{scope}
			\path[->]
			(v0) edge[above] node {$\bar{a}/{\bar{x}\bar{y}\bar{z}}$} (v2)
			(v2) edge[loop above] node[align=center] {$a/{z}$\\$\bar{a}/{z}$} (v2)
			(v0) edge[left] node {$a/{\bar{z}}$} (v1)
			(v1) edge[loop right] node[align=center]  {$a/{\bar{z}}$\\$\bar{a}/{z}$} (v1)
            (v2) edge[bend left=35,transparent] node[below] {$a/\top$} (v0) 
			;
		\end{tikzpicture}
		\subcaption{Minimal specialization of $M$}
		\label{autExSatBaseMin}
	\end{subfigure}
	\caption{Minimization example}
	\label{autExSatBaseGen}
\end{figure}
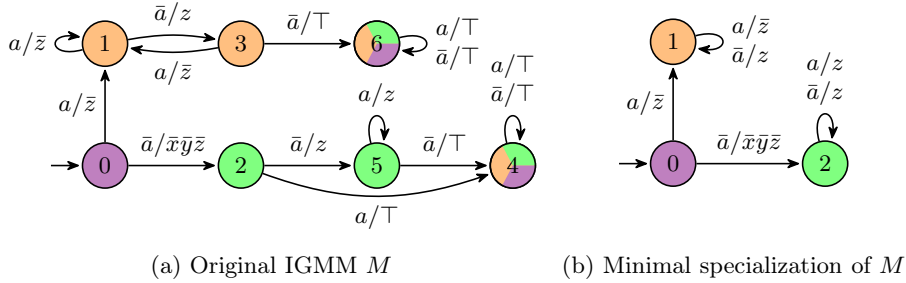

Figure~\ref{autExSatBase} illustrates this construction on an example
with a single input proposition $I=\set{a}$ (hence two input
valuations $\B^I = \set{a, \bar{a}}$), and three output propositions
$O=\set{x, y, z}$.  To simplify notations, elements of $2^{\B^O}$ are
represented as Boolean functions (happening to be cubes in this
example) rather than sets.

States have been colored to
indicate their possible membership to one of the three variational classes.
The SAT solver needs to associate each state to at least one
of them in order to satisfy the cover condition~\eqref{cond_cover},
while simultaneously respecting conditions~\eqref{cond_closure}--\eqref{cond_nonempt}.
A possible choice would be:
\textcolor{violet}{$C_0 = \{0\}$},
\textcolor{orange}{$C_1 = \{1, 3, 6\}$}, and
\textcolor{green}{$C_2 = \{2, 4, 5\}$}.
For this choice, the \textit{\textcolor{violet}{violet}} class \textcolor{violet}{$C_0$}
has only a single state, so the closure condition~\eqref{cond_closure} is trivially satisfied.
All transitions of the states in the \textit{\textcolor{orange}{orange}}
class \textcolor{orange}{$C_1$} go to states in
\textcolor{orange}{$C_1$}, also satisfying the condition.  The same
can be said of the \textit{\textcolor{green}{green}} class
\textcolor{green}{$C_2$}.

Finally, we need to check the nonempty output condition~\eqref{cond_nonempt}.
Once again, it is trivially satisfied for the
\textit{\textcolor{violet}{violet}} class \textcolor{violet}{$C_0$}.
For the \textit{\textcolor{orange}{orange}} and \textit{\textcolor{green}{green}} classes,
we need to compute their respective output.
We get $\Out(\textcolor{orange}{C_1}, a) = \bar{z}$,
$\Out(\textcolor{orange}{C_1}, \bar{a}) = z$,
$\Out(\textcolor{green}{C_2}, a) = \bar{z}$ and
$\Out(\textcolor{green}{C_2}, \bar{a}) = z$.
None of the output sets is empty, thus condition~\eqref{cond_nonempt}
is satisfied as well.
Note that, since the outgoing transitions of states 4 and 6
are self-loops compatible with all possible output valuations,
another valid choice is:
\textcolor{violet}{$C_0 = \{0, 4, 6\}$},
\textcolor{orange}{$C_1 = \{1, 3, 4, 6\}$}, and
\textcolor{green}{$C_2 = \{2, 4, 5, 6\}$}.

The corresponding specialization, constructed as described in
Theorem~\ref{theoremSATMIN}, is shown in Figure~\ref{autExSatBaseMin}.
Note that this machine is input-complete, so the incompleteness of the
specification only stems from the possible choices in the outputs.


\subsection{Proposed SAT Encoding}

We want to design an algorithm that finds a minimal specialization of a given
IGMM $M$. To do so, we will use the following approach, starting from $n = 1$:
\begin{itemize}[noitemsep,topsep=2pt]
	\item Posit that there are $n$ classes, hence,
	$n$ states in the minimal machine.
	\item Design SAT clauses ensuring cover, closure and nonempty outputs.
	\item Check if the resulting SAT problem is satisfiable.
	\item If so, construct the minimal machine described
	in Theorem~\ref{theoremSATMIN}.
	\item If not, increment $n$ by one and apply the whole process again,
	unless  $n = \left|Q\right| - 1$, which serves as a proof that the original
  machine is already minimal.
\end{itemize}

\subsubsection*{Encoding the cover and closure conditions.}
In order to guarantee that the set of classes
$S = \set{C_0, \ldots, C_{n-1}}$ satisfies both the cover and closure conditions
and that each class $C_j$ is a variation class,
we need two types of literals:

\begin{itemize}[noitemsep,topsep=2pt]
	\item $s_{q,j}$ should be true if and only if
	state $q$ belongs to the class $C_j$;
	\item $z_{i,k,j}$ should be true if
	$\Succ(C_k, i) \subseteq C_j$ for $i \in \B^I$.
\end{itemize}

The cover condition, encoded by Equation~\eqref{eq_SatPart}, guarantees that each state belongs to at least one class.\\
\begin{minipage}[t]{0.49\linewidth}
\begin{equation}
	\bigwedge_{q\in Q}\;\bigvee_{0\le j< n} s_{q,j}
  \label{eq_SatPart}
\end{equation}
\end{minipage}
\begin{minipage}[t]{0.5\linewidth}
\begin{equation}
  \bigwedge_{0\le j< n}\;\bigwedge_{\substack{q,q'\in Q\\q\nvarof q'}} \overline{s_{q,j}} \lor \overline{s_{q',j}}
  \label{eq_SatVar}
\end{equation}
\end{minipage}

Equation~\eqref{eq_SatVar} ensures that each class is a variational class:
two states $q$ and $q'$ that are not variations of each other cannot
belong to the same class.

The closure condition must ensure that for every class $C_i$ and every
input symbol $i \in \B^I$, there exists at least one class that contains all the
successor states: $\forall k, \forall i, \exists j,\ Succ(C_k, i) \subseteq C_j$.
This is expressed by the constraints~\eqref{eq_SatClos1} and~\eqref{eq_SatClos2}.
\begin{minipage}[t]{0.39\linewidth}
  \begin{equation}
    \bigwedge_{0\le k< n}\,\bigwedge_{\substack{i \in \B^I \\ \phantom{q' = \delta(q,i)}}}\,\bigvee_{0\le j < n} z_{i,k,j}
    \label{eq_SatClos1}
  \end{equation}
\end{minipage}
\begin{minipage}[t]{0.6\linewidth}
  \begin{equation}
    \bigwedge_{0\le j, k< n}\;\bigwedge_{\substack{q, q' \in Q, i \in \B^I \\ q' = \delta(q,i)}} (z_{i,k,j} \land s_{q,k}) \rightarrow s_{q',j}
    \label{eq_SatClos2}
  \end{equation}
\end{minipage}
The constraint~\eqref{eq_SatClos1} ensures that at least one $C_j$ contains
$\Succ(C_k, i)$, while~\eqref{eq_SatClos2} ensures this mapping of classes
matches the transitions of $M$.

\subsubsection*{Encoding the nonempty output condition.}
\label{encNonEmpty}

Each class in $S$ being a variation class is necessary
but not sufficient to satisfy the nonempty output condition.
We indeed want to guarantee that for any input $i$,
all states in a given class can agree on at least one common output valuation.

However it is possible to have three or more states (like
    \begin{tikzpicture}[smallautomaton,baseline=(s.base)]
      \node[state] (s) {\small $0$};
      \path[->] (s) edge[loop right] node[inner sep=0pt](x){$\,a/\{xy,x\bar y\}$} (s);
      \path[use as bounding box] (s.north west) rectangle(x.south east);
    \end{tikzpicture},
    \begin{tikzpicture}[smallautomaton,baseline=(s.base)]
      \node[state] (s) {\small $1$};
      \path[->] (s) edge[loop right] node[inner sep=0pt](x){$\,a/\{\bar xy,x\bar y\}$} (s);
      \path[use as bounding box] (s.north west) rectangle(x.south east);
    \end{tikzpicture}, and
    \begin{tikzpicture}[smallautomaton,baseline=(s.base)]
      \node[state] (s) {\small $2$};
      \path[->] (s) edge[loop right] node[inner sep=0pt](x){$\,a/\{xy,\bar xy\}$} (s);
      \path[use as bounding box] (s.north west) rectangle(x.south east);
    \end{tikzpicture})
    that are all variations of one another, but still cannot
    agree on a common output.

    This situation cannot occur in \textsc{MeMin} since their model uses
    \emph{cubes} as outputs rather than arbitrary sets of valuations
    as in our model.
    A useful property of cubes is that if the
    pairwise intersections of all cubes in a set are nonempty, then
    the intersection of all cubes in the set is necessarily nonempty as well.


Since \emph{cubes} are not expressive enough for our model,
we will therefore generalize the output
as discussed earlier in Section \ref{secCompToReg}:
we represent the arbitrary set of valuations produced by the output
function $\lambda$ as a set of cubes whose disjunction yields the original set.
For $q \in Q$ and $i \in \B^I$, we partition the set of valuations
$\lambda(q, i)$ into cubes, relying on the~\citet{minato.92.sasimi} algorithm,
and denote the obtained set of cubes as $\mathrm{CS}(\lambda(q, i))$.

Our approach for ensuring that there exists a common output is to
search for disjoint cubes and exclude them from the possible outputs
by selectively deactivating them if necessary; an active cube is a set
in which we will be looking for an output valuation that the whole
class can agree on. To express this, we need two new types of
literals:
\begin{itemize}[noitemsep,topsep=2pt]
	\item $a_{c,q,i}$ should be true iff
          the particular instance of the cube $c\in \mathrm{CS}(\lambda(q,i))$ used
          in the output of state $q$ when reading $i$ is \emph{active};
	\item $\mathit{sc}_{q,q'}$ should be true iff
	$\exists C_j \in S$ such that $q\in C_j$ and $q'\in C_j$
      \end{itemize}
The selective deactivation of a cube can then be expressed by the following:

\begin{minipage}[t]{.46\textwidth}
\begin{equation}
  \bigwedge_{\substack{q, q' \in Q \\0 \le j < n}} (s_{q, j} \land s_{q', j})
  \rightarrow \mathit{sc}_{q,q'}
  \label{eq_SatSameClass}
\end{equation}
\end{minipage}
\hfill
\begin{minipage}[t]{.46\textwidth}
\begin{equation}
\bigwedge_{\substack{q \in Q,\, i \in \B^I \\\delta(q, i) \text{~is defined} }}\!\!\bigvee_{{c}\in \mathrm{CS}(\lambda(q, i))} a_{c,q,i}\\
\label{eq_SatNEPart}
\end{equation}
\end{minipage}

\begin{equation}
\bigwedge_{\substack{q, q' \in Q,\, i \in \B^I \\
    \delta(q, i) \text{~is defined}\\\delta(q', i) \text{~is defined} }}
\;
  \bigwedge_{\substack{c\in \mathrm{CS}(\lambda(q, i)) \\
  c'\in \mathrm{CS}(\lambda(q', i)) \\
  c \cap c' = \emptyset}}
(a_{c,q,i} \land a_{c',q',i}) \rightarrow \overline{\mathit{sc}_{q,q'}}.
  \label{eq_SatDeact}
\end{equation}

Constraint~\eqref{eq_SatSameClass} ensures that $\mathit{sc}_{q,q'}$ is true if there exists
a class containing both $q$ and $q'$, in accordance with the expected definition.

Constraint~\eqref{eq_SatNEPart} guarantees that at least one of the cubes
in the output $\lambda(q, i)$ is active,
causing the restricted output to be nonempty.

Constraint~\eqref{eq_SatDeact} expresses selective deactivation and only needs to be
added for a given $q, q' \in Q$ and $i \in \B^I$ if
$\delta(q, i)$ and $\delta(q', i)$ are properly defined.
This formula guarantees that if there exists a class to which $q$ and $q'$
belong to (i.e., $\mathit{sc}_{q,q'}$ is true) but there also exist disjoint cubes
in the partition of their respective outputs, then
we deactivate at least one of these:
only cubes that intersect can be both activated.
Thus, this constraint guarantees the nonempty output condition.

Since encoding an output set requires a number of cubes exponential
in $|O|$, the above encoding uses
$\mathrm{O}(|Q|(2^{|I|+|O|}+|Q|)+n^2 \cdot 2^{|I|})$ variables as well
as
$\mathrm{O}(Q^2(n+2^{2|O|})+n^2 \cdot 2^{|I|}+|\delta|(2^{|O|}+n^2))$
clauses.  We use additional optimizations to limit the number of
clauses, and make the algorithm more practical despite its frightening
theoretical worst case.  In particular the CEGAR approach of
Section~\ref{sec:cegar} strives to avoid introducing constraints
\eqref{eq_SatSameClass}--\eqref{eq_SatDeact}.

\subsection{Adjustment of Prior Optimizations}

Constructing the SAT problem iteratively starting from $n = 1$
would be grossly inefficient.
We can instead notice that two states that are not variations of each
other can never be in the same class.
Thus, assuming we can find $k$ states that are not pairwise variations
of one another, we can infer that we need at least as many classes
as there are states in this set, providing a lower bound for $n$.
This idea was first introduced in~\citet{abel.15.iccad};
however, performing a more careful inspection of the constraints
with respect to this ``partial solution'' allows us
to reduce the number of constraints and literals needed.

The nonemptiness condition involves the creation of many literals and clauses
and necessitates an expensive preprocessing step to decompose the
arbitrary output sets returned by output function
($\lambda \colon \left(Q, \B^{I}\right) \rightarrow 2^{\B^{O}}\setminus \{\emptyset\}$)
into disjunctions of cubes ($\lambda \colon \left(Q, \B^{I}\right) \rightarrow 2^{\K^{O}}\setminus \{\emptyset\}$).
We avoid adding
unnecessary nonempty output clauses in a counter-example guided fashion.
Violation of these conditions can easily be detected
before constructing the minimized machine.
If detected, a small set of these constraints is added to SAT problem
excluding this particular violation.
In many cases, this optimization greatly reduces the number of
literals and constraints needed, to the extent we can often
avoid their use altogether.

From now on, we consider an IGMM with $N$ states
$Q=\set{q_0, q_1, \ldots, q_{N-1}}$.

\subsubsection*{Variation matrix.}
We first need to determine which states are not pairwise
variations of one another in order to extract a partial solution
and perform simplifications on the constraints.
We will compute a square matrix of size $N\times N$ called $\mathrm{mat}$
such that $\mathrm{mat}[k][\ell] = 1$ if and only if $q_k\nvarof q_\ell$
in the following fashion:
\begin{enumerate}
  \item Initialize all entries of $\mathrm{mat}$ to $0$.
  \item Iterate over all pairs $\tup{k, \ell}$ with $0 \le k < \ell < N$.
  If the entry $\mathrm{mat}[k][\ell]$ is $0$, check if $\exists i \in \B^I$
  such that $\lambda(q_k, i) \cap \lambda(q_l, i) = \emptyset$. If it exists,
  $\mathrm{mat}[k][\ell] \gets 1$.
  \item For all pairs $\tup{k, \ell}$ whose associated value $\mathrm{mat}[k][\ell]$
  changed from $0$ to $1$, set all existing predecessor pairs
  $\tup{m,n}$ with $m< n$ under the same input to $1$ as well, that is,
  $\exists i \in \B^I$ such that $\delta(q_m, i) = q_k$ and
  $\delta(q_n, n) = q_l$. Note that we may need to propagate these changes
  to the predecessors of $\tup{m, n}$.
\end{enumerate}

As ``being a variation of'' is a symmetric, reflexive relation, we only
compute the elements above the main diagonal of the matrix.
The intuition behind this algorithm is that two states $q$ and $q'$ are
not variations of one another if either:
\begin{itemize}
	\item There exists an input symbol for which the output sets are disjoint.
	\item There exists a pair of states which are not variations of one another
	and that can be reached from $q$ and $q'$ under the same input
	sequence.
\end{itemize}

The complexity of this algorithm is
$\mathrm{O}(|Q|^2 \cdot 2^{|I|})$
if we assume that the disjointness of the output sets can be checked in
constant time; see~\citet{abel.15.iccad}.
This assumption is not correct in general: testing disjointness
for cubes has a complexity linear in the number of input propositions.
On the other hand, testing disjointness for generalized Mealy machines
that use arbitrary sets of valuations has a complexity exponential in the
number of input propositions. This increased complexity is however
counterbalanced by the succinctness the use of arbitrary sets allows.

As an example, given $2m$ output propositions $o_0, \ldots, o_{2m-1}$,
consider the set of output valuations expressed as a disjunction of cubes
$\bigvee_{0 \le k < m} o_{2k}\,\overline{o_{2k+1}} \lor
\overline{o_{2k}}\, o_{2k+1}$. Exponentially many \emph{disjoint} cubes are
needed to represent this set. Thus, a non-deterministic Mealy machine
labeled by output cubes will incur an exponential number of computations
performed in linear time, whereas a generalized Mealy machine
will only perform a single test with
exponential runtime.


\subsubsection*{Computing a partial solution.}
The partial solution corresponds to a set of states such that none of them is
a variation of any other state in the set.
Thus, none of these states can belong to the same (variation) class.
The size of this set is therefore a lower bound for the number of states in
the minimal machine.

Finding the largest partial solution is an NP-hard problem; we therefore
use the greedy heuristic described in~\citet{abel.15.iccad}.
For each state $q$ of $M$, we count the number of states $q'$ such that
$q$ is not a variation of $q'$; call this number $\mathit{nvc}_q$.
We then successively add to the partial solution the states
that have the highest $\mathit{nvc}_q$ but are not variations
of any state already inserted.

\subsubsection*{CEGAR approach to ensure the nonempty output condition.}\label{sec:cegar}
Assuming a solution satisfying the cover and closure constraints has already
been found, we then need to check if
said solution satisfies the nonempty output condition.
If this is indeed the case, we can then construct and return a minimal machine.

If the condition is not satisfied, we look for one or
more combinations of classes and input symbols such that
$\Succ(C_k, i) = \emptyset$.
We add for the states in $C_k$ and the input symbol $i$
the constraints described in Section~\ref{encNonEmpty}, and for these states
and input symbols only. Then we check if the problem is still satisfiable.

If it is not, then we need to increase the number of classes to find
a valid solution.  If it is, the solution either respects
condition~\eqref{cond_nonempt} and we can return a minimal machine, or
it does not and the process of selectively adding constraints is
repeated.  Either way, this \emph{counter-example guided abstraction
  refinement} (CEGAR) scheme ensures termination, as the problem is
either shown to be unsatisfiable or solved through iterative exclusion
of all violations of condition~\eqref{cond_nonempt}.

\subsection{Algorithm}

The optimizations described previously yield Algorithm~\ref{algoSAT1}.

\begin{algorithm}[t]
  \KwData{a machine $M=\tup{I,O,Q,q_{\mathit{init}},\delta,\lambda}$}
  \KwResult{a minimal specialization $M'$}
  \tcc{Computing the variation matrix}
  bool[][] $\mathrm{mat}$ $\gets$ isNotVariationOf($M$)\;
  \tcc{Looking for a partial solution P}
  set $P \gets$ extractPartialSol($\mathrm{mat}$)\;
  clauses $\gets$ empty list\;
  \tcc{Using the lower bound inferred from P}
  \For{$n\gets \left|P\right| \KwTo \left|Q\right|-1$}{
    addCoverCondition(clauses, $M$, $P$, $\mathrm{mat}$, $n$)\;
    addClosureCondition(clauses, $M$, $P$, $\mathrm{mat}$, $n$)\;
    \tcc{Solving the cover and closure conditions}
    (sat, solution) $\gets$ satSolver(clauses)\;
    \While{sat}{
      \If{verifyNonEmpty($M$, solution)}{
        \KwRet buildMachine($M$, solution)\;
      }
      \tcc{Adding the relevant nonemptiness clauses}
      addNonemptinessCondition(clauses, $M$, solution)\;
      (sat, solution) $\gets$ satSolver(clauses)\;
    }
  }
  \tcc{If no solution has been found, return M}
  \KwRet copyMachine($M$)\;
  \caption{SAT-based minimization}
  \label{algoSAT1}
\end{algorithm}

\subsubsection*{Further optimizations and comparison to \textsc{MeMin}.}
The proposed algorithm relies on the general approach outline
in~\citet{abel.15.iccad}, as well as the SAT encoding for the cover and closure
conditions.
We find a partial solution by using a similar heuristic and adapt some
optimizations found in their source code, which are neither detailed
in their paper nor here due to a lack of space.

The main difference lies in the increased expressiveness of the input and output
symbols that causes some significant changes.
In particular, we added the nonemptiness condition to guarantee
correctness, as well as a CEGAR-based implementation to maintain performance.
Other improvements mainly stem from a better usage of the partial solution.

For instance, each state $q$ of the partial solution is associated to
``its own'' class $C_j$. Since the matching literal $s_{q,j}$ is trivially true,
it can be omitted by replacing all its occurrences by true.
States belonging to the partial solution have other peculiarities that
can be leveraged to reduce the number of possible successor classes,
further reducing the amount of literals and clauses needed.

We therefore require fewer literals and clauses, trading
a more complex construction of the SAT problem
for a reduced memory footprint.
The impact of these improvements is detailed in Section~\ref{secBench}.

The Mealy machine described by~\citet{abel.15.iccad} come in two flavors:
One with an explicit initial state and a second one where all states are
considered to be possible initial states.
While our approach does
explicit an initial state, it does not further influence the resulting minimal machine
when all original states are reachable.

\section{Bisimulation with Output Assignment}
\label{secBisim}

We introduce in this section another approach tailored to our
primary use case, that is, efficient reduction of control strategies in the
context of reactive synthesis. This technique, based on the $\specof$
specialization relation, yields non-minimal but ``relatively small'' machines at
significantly reduced runtimes.

Given two states $q$ and $q'$ such that $q'\specof q$, one idea is to
restrict the possible outputs of $q$ to match those of $q'$.
Concretely, for all inputs $i\in \B^I$, we restrict $\lambda(q,i)$ to
its subset $\lambda(q',i)$; $q$ and $q'$ thus become
bisimilar, allowing us to merge them. In practice, rather than restricting
the output first then reducing bisimilar states to their quotient,
 we instead directly build a machine that is minimal
 with respect to $\specof$ where all transitions going to $q$
 are redirected to $q'$.

Note that if two states $q$ and $q'$ are bisimilar,
then necessarily $q'\specof q$ and $q\specof q'$: therefore, both states will be
merged by our approach. As a consequence, the resulting machine is always
smaller than the bisimulation quotient of the original machine
(as shown in Section~\ref{secBench}).

\subsection{Reducing Machines with $\specof$}

Our algorithm builds upon the following theorem:
\begin{theorem}
	\label{theoremSpecReduc}
	Let $M = \tup{I, O, Q, q_{\mathit{init}}, \delta, \lambda}$ be
        an IGMM, and $r\colon Q\to Q$ be a mapping satisfying
        $r(q)\specof q$. Define
        $M' = \tup{I, O, Q', q_{\mathit{init}}', \delta', \lambda}$ as
        an IGMM where $Q' = \mathit{r(Q)}$,
        $q'_{\mathit{init}}=r(q_{\mathit{init}})$ and
        $\delta' (q, i) = r(\delta (q, i))$ for all states $q$ and input $i$.
        Then $M'$ is a specialization of $M$.
\end{theorem}


Intuitively, if a state $q$ is remapped to a
state $q'\specof q$, then the set of words $w$ that can be output for an
input $i$ is simply reduced to a subset of the original output.
The smaller the image $r(Q)$, the more significant the reduction performed on
the machine. Thus, to find a suitable function $r$,
we map each state $q$ to one of the
\emph{minimal elements} of the $\specof$ preorder, also called
the \emph{representative states}.

\begin{figure}[bt]
  \begin{minipage}{.35\textwidth}
  \centering
  \begin{tikzpicture}[mediumautomaton, yscale=1.164
    ]
    \node at (0,2) (n46) {$\{4, 6\}$};
    \node at (0,1) (n3) {$\{3\}$};
    \node at (-1.5,0) (n2) {$\{2\}$};
    \node at (-.5,0) (n0) {$\{0\}$};
    \node at (.5,0) (n1) {$\{1\}$};
    \node at (1.5,0) (n5) {$\{5\}$};
    \draw [->] (n46) edge[bend right] (n2);
    \draw [->] (n46) edge[bend right=15] (n0);
    \draw [->] (n46) -- (n3);
    \draw [->] (n46) edge[bend left=15] (n1);
    \draw [->] (n46) edge[bend left] (n5);
    \draw [->] (n3) edge[bend right=5] (n0);
    \draw [->] (n3) edge[bend left=5] (n1);
    \draw[dashed,thin,rounded corners=1mm] ($(n2.north west)+(-2mm,2mm)$) rectangle ($(n5.south east)+ (2mm,-2mm)$);
    \node[below=-1.5mm] at (n2.south -| n3) {leaves};
  \end{tikzpicture}
  \caption{Specialization graph of the IGMM of Fig.~\ref{autExSatBase}}
  \label{fig:graph_ex}
  \end{minipage}
  \hfill
  \begin{minipage}{.3\textwidth}
    \centering
    \begin{tabular}{lcl}
      $q$ && $\mathllap{r(}q\mathrlap{)}$ \\
      \midrule
      0 &$\to$ & 0 \\
      1 &$\to$ & 1 \\
      2 &$\to$ & 2 \\
      3 &$\to$ & 1 \\
      4 &$\to$ & 1 \\
      5 &$\to$ & 5 \\
      6 &$\to$ & 1 \\
    \end{tabular}
    \vspace*{-1mm}
    \caption{Chosen representative mapping.\label{fig:autMapBisim}}
  \end{minipage}
  \hfill
  \begin{minipage}{.33\textwidth}
  \centering
  \begin{tikzpicture}[mediumautomaton,node distance=.8cm and 1cm]
    \begin{scope}[local bounding box=aut]
      \node[initial,lstate, fill=violet!50] (v0) {$0$};
      \node[lstate,above=of v0, fill=orange!50] (v1) {$1$};
      \node[lstate,right=of v0, fill=green!50] (v2) {$2$};
      \node[lstate,above=of v2, fill=green!50] (v5) {$5$};
    \end{scope}
    \path[->]
    (v0) edge[left] node {${a}/{\bar{z}}$} (v1)
    (v0) edge[below] node {${\bar{a}}/{\bar{x}\bar{y}\bar{z}}$} (v2)
    (v1) edge[loop left] node {${a}/{\bar{z}}$} (v1)
    (v1) edge[loop above] node {${\bar{a}}/{z}$} (v3)
    (v2) edge node[above right=-3pt] {${a}/\top$} (v1)
    (v2) edge[right] node {${\bar{a}}/{z}$} (v5)
    (v5) edge[above] node {${\bar{a}}/\top$} (v1)
    (v5) edge[loop above] node {${a}/{z}$} (v5)
    ;
  \end{tikzpicture}
  \caption{IGMM obtained by reducing that of Fig.~\ref{autExSatBase}\label{autExBisim}}
  \end{minipage}
\end{figure}

\begin{definition}[Specialization graph]
  A \emph{specialization graph} of an IGMM
  $M = \tup{I, O, Q, q_{\mathit{init}}, \delta, \lambda}$ is the
  \emph{condensation graph} of the directed graph representing the
  relation $\specof$: the vertices of the specialization graph
  are sets that form a partition of $Q$ such that two states $q$ and
  $q'$ belong to the same vertex if $q \specof q'$ and $q' \specof q$;
  there is an edge
  $\{q_1,q_2,...\} \longrightarrow \{q'_1,q'_2,...\}$ if and only if
  $q'_i \specof q_j$ for some (or equivalently all) $i,j$.  Note that
  this graph is necessarily acyclic.
\end{definition}

Fig.~\ref{fig:graph_ex} shows the specialization graph associated to
the machine of Fig.~\ref{autExSatBase}.

\begin{definition}[Representative of a state]
  Given two states $q$ and $q'$ of an IGMM, $q'$ is
  a \emph{representative} of $q$ if, in the specialization graph of $M$, $q'$
  belongs to a leaf that can be reached from the vertex containing $q$.
  In other words, $q'$ is a representative of $q$ if $q' \specof q$ and $q'$ is
  a minimal element of the $\specof$ preorder.
\end{definition}

Note that any state has at least one representative.  In
Fig.~\ref{fig:graph_ex} we see that $0$ represents $0$,
$3$, $4$, and $6$.  States $3$, $4$, and $6$ can be represented by
$0$ or $1$.

By picking one state in each leaf, we obtain a set of
representative states that cover all states of the IGMM.  We then
apply Theorem~\ref{theoremSpecReduc} to a function $r$ that maps
each state to its representative in this cover.  In Fig.~\ref{fig:graph_ex},
all leaves are singletons, so the set
$\{0,1,2,5\}$ contains representatives for all states. Applying
Th.~\ref{theoremSpecReduc} using $r$ from Fig.~\ref{fig:autMapBisim}
yields the machine shown in Fig.~\ref{autExBisim}.  Note that while this
machine is smaller than the original, it is still bigger than the
minimal machine of Fig.~\ref{autExSatBaseMin}, as this approach does not
appraise the variation relation $\varof$.

\subsection{Implementing $\specof$}

We now introduce an effective decision procedure for $q \specof q'$.
Note that $\specof$ can be defined recursively like a simulation
relation.  Assuming, without loss of generality, that the IGMM is
input-complete, $\specof$ is the coarsest relation satisfying:
\[
  q'\specof q \Longrightarrow \forall i\in \B^I, \begin{cases}
    \lambda(q',i) \subseteq \lambda(q,i) \\
    \delta(q',i) \specof \delta(q,i) \\
  \end{cases}
\]
As a consequence, $\specof$ can be decided using any technique that is suitable
for computing simulation
relations~\citet{henzinger.95.focs,etessami.00.concur}.
Our implementation relies on a straightforward adaptation of the technique
of signatures described by~\citet[][Sec.~4.2]{babiak.13.spin}: for
each state $q$, we compute its \emph{signature} $\mathrm{sig}(q)$, that
is, a Boolean formula (represented as a BDD) encoding the outgoing
transitions of that state such that
$\mathrm{sig}(q) \Rightarrow \mathrm{sig}(q')$ if and only if $q\specof q'$.
Using these signatures, it becomes easy to build the
\emph{specialization graph} and derive a remapping function $r$.

Note that, even if $\specof$ can be computed like a simulation, we do
not use it to build a bisimulation quotient.  The remapping applied in
Th.~\ref{theoremSpecReduc} does not correspond to the quotient of $M$ by
the equivalence relation induced by $\specof$.


\section{Benchmarks}\label{secBench}

The two approaches described in Sections \ref{secMin} and \ref{secBisim}
have been implemented within Spot
2.10~\citet{duret.16.atva2}, a toolbox for $\omega$-automata
manipulation, and used in our SyntComp'21
submission~\citet{renkin.21.synt}.  The following benchmarks
are based on a development version of Spot\footnote{For instructions
  to reproduce, see \url{https://www.lrde.epita.fr/~philipp/forte22/}}
that features efficient  variation checks (verifying whether $q\varof q'$)
thanks to an improved representation of cubes.

We benchmark the two proposed approaches against \textsc{MeMin},
against a simple bisimulation-based approach, and against one another.
The \textsc{MeMin} tool has already been shown~\citet{abel.15.iccad} to
be superior to existing tools like
\textsc{Bica}~\citet{pena.99.cadics},
\textsc{Stamina}~\citet{rho.94.cadics}, and
\textsc{Cosme}~\citet{alberto.13.ocs}; we are not aware of more recent
contenders.  For this reason, we only compare our approaches to
\textsc{MeMin}.

In a similar manner to~\citet{abel.15.iccad}, we use the ISM
benchmarks~\citet{kam1994fully} as well as the MCNC benchmark
suite~\citet{yang1991logic}.  These benchmarks share a severe drawback: they
only feature very small instances.  \textsc{MeMin} is able to solve any of
these instances in less than a second.  We therefore extend the set of
benchmarks with our main use-cases: Mealy machines corresponding to control
strategies obtained from SYNTCOMP LTL specifications~\citet{jacobs20205th}.

As mentioned in Section \ref{secCompToReg}, \textsc{MeMin} processes Mealy
machines, encoded using the the KISS2 input format~\citet{yang1991logic},
whose output can be chosen from a cube. However, the IGMM formalism we promote
allows an arbitrary set of output valuations instead.
This is particularly relevant for the SYNTCOMP benchmark, as the LTL
specifications from which the sample's Mealy machines are derived often fail to fully
specify the output. In order to (1) show the benefits of the generalized
formalism while (2) still allowing comparisons with \textsc{MeMin}, we
prepared two versions of each SYNTCOMP input: the ``full'' version features
arbitrary sets of output valuations that cannot be processed by \textsc{MeMin},
while in the ``cube'' version said sets have been
replaced by the first cube produced by the Minato algorithm~\citet{minato.92.sasimi}
on the original output set. The ACM and MCNC benchmarks, on the other hand,
already use a single output cube in the first place.

\begin{figure}[t]
  \begin{minipage}[t]{0.48\textwidth}
    \centering
      \resizebox {1.\textwidth} {!}
      {
        \input{tot_time.tex}
      }
      \caption{Log-log plot of runtimes. The legend $a/b$ stands for
        $a$ cases above diagonal, and $b$ below.}
    \label{fig_tottime}
  \end{minipage}
  \hfill
  \begin{minipage}[t]{0.48\linewidth}
    \centering
      \resizebox {1.\textwidth} {!}
      {
        \input{n_lit_clause.tex}
      }
    \caption{Comparison of the number of literals and clauses in the encodings.}
    \label{fig_nclauses}
  \end{minipage}
\end{figure}

\begin{table}[t]
  \begin{center}
    \begin{tabular}{lccccc|ccc@{~}ccc|ccc@{~}cc}
      &      &      &      &  & & & \multicolumn{2}{c}{$\frac{\mathit{size}}{\mathit{orig}}$} & \multicolumn{2}{c}{$\frac{\mathit{size}}{\mathit{min}}$} &      &                  & \multicolumn{2}{c}{$\frac{\mathit{size}}{\mathit{orig}}$} & \multicolumn{2}{c}{$\frac{\mathit{size}}{\mathit{min}}$}            \\

                                                 & >(1) & >(2) & >(3)          & >(4)          &  &  & avg.              & md. & avg.           & md.              &  &  & avg.              & md.            & avg.            & md.                         \\
      \cmidrule{1-5} \cmidrule{8-11} \cmidrule{14-17}

                                original &                        114 &           304 &           271 &           314 &  &  & 1.00 & 1.0 & 6.56 &              1.0 &  &  &              1.00 & 1.00 &           12.23 &              1.77 \\
                        (1) bisim (full) &              \phantom{000} &           249 &           214 &           275 &  &  & 0.94 & 1.0 & 1.85 &              1.0 &  &  &              0.88 & 1.00 & \phantom{0}2.72 &              1.50 \\
         (2) bisi\rlap{m w/ o.a. (full)} &              \phantom{000} & \phantom{00}0 & \phantom{0}68 & \phantom{0}84 &  &  & 0.83 & 1.0 & 1.55 &              1.0 &  &  &              0.66 & 0.67 & \phantom{0}2.10 &              1.00 \\
(3) \textsc{MeMin} (minima\rlap{l cube)} &              \phantom{000} & \phantom{0}74 & \phantom{00}0 & \phantom{0}77 &  &  & 0.81 & 1.0 & 1.13 &              1.0 &  &  &              0.63 & 0.69 & \phantom{0}1.27 &              1.00 \\
                          (4) SAT (full) & \tikzmark{b1}\phantom{000} & \phantom{00}0 & \phantom{00}0 & \phantom{00}0 &  &  & 0.77 & 1.0 & 1.00 & 1.0\tikzmark{b2} &  &  & \tikzmark{b3}0.54 & 0.56 & \phantom{0}1.00 & 1.00\tikzmark{b4} \\

      \\[1em]
    \end{tabular}
  \end{center}
  \begin{tikzpicture}[overlay,remember picture]
    \draw [decoration={brace},decorate,thick]
        ($(b2)+(0,-2mm)$) -- node [below=2pt,align=center] {all 634 instances\\without timeout} ($(b1)+(0,-2mm)$) ;
    \draw [decoration={brace},decorate,thick]
        ($(b4)+(0,-2mm)$) -- node [below=2pt,align=center] {314 non-minimal\\\llap{instan}ces without timeout} ($(b3)+(0,-2mm)$) ;
  \end{tikzpicture}%
  \caption{Statistics about our three reduction algorithms. The leftmost pane
    counts the number of instances where algorithm (y) yields a smaller result than
    algorithm (x); as an example,
    bisimulation with output assignment (2) outperforms
    standard bisimulation (1) in 249 cases. The middle pane presents mean
    (avg.) and median (md.) size ratios relative to the
    original size and the minimal size of the sample machines.
    The rightmost pane presents similar statistics while ignoring
    all instances that were already minimal in the first place.\label{tab:stats}}
\end{table}

Figure~\ref{fig_tottime} displays a log-log plot comparing our different methods to
\textsc{MeMin}, using only the ``cube'' instances.\footnote{A 30 minute
  timeout was enforced for all instances. The benchmarks were run on an
  Asus G14 with a Ryzen 4800HS CPU with 16GB of RAM and no swap }. The
label ``\emph{bisim. w/ o.a.}''  refers to the approach outlined in
Section~\ref{secBisim}, ``\emph{bisim.}'', to a simple bisimulation
quotient, and ``\emph{SAT}'', to the approach of Section~\ref{secMin}.
Points on the black diagonal stand for cases where \textsc{MeMin} and the
method being tested had equal runtime; cases above this line favor
\textsc{MeMin}, while cases below favor the aforementioned methods.  
Points on the dotted line at the edges of the figure represent timeouts.
Only \textsc{MeMin} fails this way, on 10 instances.
Figure~\ref{fig_nclauses} compares the maximal number of literals and clauses
used to perform the SAT-based minimization by \textsc{MeMin} or by our
implementation.  These two figures only describe ``cube'' instances, as
\textsc{MeMin} needs to be able to process the sample machines.

To study the benefits of our IGMM model's generic outputs, Table~\ref{tab:stats}
compares the relative reduction ratios achieved by the various methods
w.r.t. other methods as well as the original and minimal size of the sample
machines.  We use the ``full'' inputs everywhere with the exception of
\textsc{MeMin}.

\subsubsection{Interpretation.}
Reduction via bisimulation solves all instances and has been proven to be by far the
fastest method (Fig.~\ref{fig_tottime}), but also the coarsest, with
a mere $0.94$ reduction ratio (Tab.\ref{tab:stats}).
Bisimulation with output assignment achieves a better reduction ratio of $0.83$, very
close to \textsc{MeMin}'s $0.81$.

In most cases, the proposed SAT-based approaches remain significantly slower than
approaches based on bisimulation (Fig.~\ref{fig_tottime}). Our SAT-based
algorithm is sometimes slower than \textsc{MeMin}'s, as the model's increased
expressiveness requires a more complex method. However,
improving the use of partial solutions and increasing the expressiveness
of the input symbols significantly reduce the size of the encoding of
the intermediate SAT problems featured in our method (Fig.~\ref{fig_nclauses}),
hence, achieve a lower memory footprint.
Points on the horizontal line at the bottom of Figure~\ref{fig_nclauses}
correspond to instances that have already been proven minimal,
since the partial solution is equal to the entire set of states:
in these cases, no further reduction is required.

Finally, the increased expressiveness of our model results in
significantly smaller minimal machines, as shown by the $1.27$
reduction ratio of \textsc{MeMin}'s cube-based machines compared to
the minimisation of generic IGMMs derived from the same specification.
There are also 74 cases where this superior expressiveness allows the
bisimulation with output assignment to beat \textsc{MeMin}.

\section{Conclusion}\label{secConcl}

We introduced a generalized model for incompletely specified Mealy
machines, whose output is an arbitrary choice between multiple
possible valuations.
We have presented two reduction techniques on this model,
and compared them against the state-of-the-art minimization tool
\textsc{MeMin} (where the output choices are restricted to a cube).

The first technique is a SAT-based approach inspired by {\sc
  MeMin}~\citet{abel.15.iccad} that yields a minimal machine. Thanks to
this generalized model and an improved use of the partial solution, we
use substantially fewer clauses and literals.

The second technique yields a reduced yet not necessarily minimal
machine by relying on the notion of state specialization.  Compared
to the SAT-based approach, this technique offers a good compromise
between the time spent performing the reduction, and the actual
state-space reduction, especially for the cases derived from SYNTCOMP
from which our initial motivation originated.

Both techniques are implemented in Spot 2.10.  They have been used in
our entry to the 2021 Synthesis Competition~\citet{renkin.21.synt}.
Spot comes with Python bindings that make it possible to experiment
with these techniques and compare their respective effects\footnote{See: \url{https://spot.lrde.epita.fr/ipynb/synthesis.html}.}.

\bibliographystyle{abbrvnat}
\bibliography{mc}

\ifappendix\else
\end{document}